\documentclass[aps,reprint,prl,amsmath,amssymb,citeautoscript,numerical,superscriptaddress,twocolumn,showpacs]{revtex4-1}
\usepackage{graphicx}
\usepackage{amsmath}
\usepackage{bm}

\newcommand{\p}{\partial}

\newcommand{\vF}{v_{\text{F}}}

\begin{document}


\title{Giant Plasmon Instability in 
  Dual-Grating-Gate Graphene Field-Effect Transistor}

\author{Y. Koseki}
\author{V. Ryzhii}
\author{T. Otsuji}
\affiliation{Research Institute of Electrical Communication,
  Tohoku University, Sendai 980-8577, Japan}
\author{V. V. Popov}
\affiliation{Kotelnikov Institute of Radio Engineering and 
  Electronics (Saratov Branch), 410019 Saratov, Russia}
\affiliation{Saratov State University, Saratov 410012, Russia}
\affiliation{Saratov Scientiﬁc Center of the Russian Academy of 
Sciences, Saratov 410028, Russia}
\author{A. Satou}\email{a-satou@riec.tohoku.ac.jp}
\affiliation{Research Institute of Electrical Communication,
  Tohoku University, Sendai 980-8577, Japan}
\date{\today}

\begin{abstract}
We study instability of plasmons in a dual-grating-gate graphene
field-effect transistor induced by dc current injection 
using self-consistent
simulations with the Boltzmann equation. With only the
acoustic-phonon-limited electron scattering, it is demonstrated that
a total growth rate of the plasmon instability,
with the terahertz/mid-infrared range of the frequency,
can exceed $4\times10^{12}$ s$^{-1}$ at room temperature, 
which is an order of magnitude larger than in two-dimensional 
electron gases based on usual semiconductors.
By comparing the simulation results with existing theory,
it is revealed that the giant total growth rate originates from
simultaneous occurence of
the so-called Dyakonov-Shur and Ryzhii-Satou-Shur instabilities.
\end{abstract}

\maketitle



Electronic, hydrodynamic, and electromagnetic properties of 
two-dimensional (2D) plasmons in channels of field-effect transistors
(FETs) have been 
investigated extensively for their utilization to terahertz (THz) 
devices~\cite{Allen1977,Dyakonov1993,Dyakonov1996,Veksler2006,
Dyer2013,Rozhansky2015}
(see also review 
papers~\cite{Shur2003,Knap2009,Otsuji2014a,Otsuji2015a}
and references therein).
Especially, plasmon instability is one of the most important 
properties to realize compact, room-temperature operating THz 
sources. Self-excitation of plasmons due to instability induces 
ac voltages in the gate electrodes and, in turn, leads to 
the emission of THz waves.

\begin{figure}[t]
  \begin{center}
    \includegraphics[width=8.8cm]{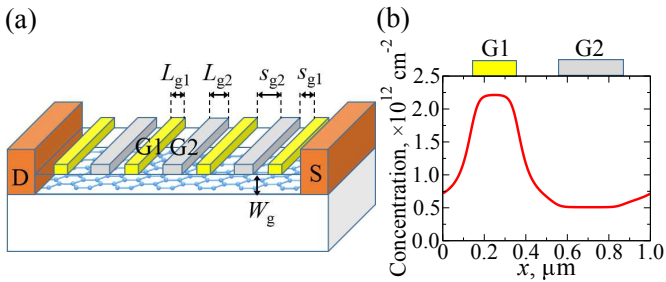}
  \caption{\label{FigSchematics}
    Schematic views of (a) a dual-grating-gate graphene FET
    and (b) a profile of steady-state electron concentration
    under consideration.}
  \end{center}
\end{figure}

The so-called Dyakonov-Shur (DS) 
instability~\cite{Dyakonov1993,Dmitriev1997,Dyakonov2005,Popov2008} 
and Ryzhii-Satou-Shur (RSS) 
instability~\cite{Ryzhii2005,Ryzhii2006e,Ryzhii2008c} 
in single-gate FETs were proposed theoretically
as mechanisms of plasmon instability by dc current injection
through the transistor channel.
The DS instability originates from the Doppler shift effect
at asymmetric boundaries in the channel,
i.e., zero time-variation of pontential at the source contact and 
zero time-variation of electron velocity near the drain contact,
which are naturally realized by operating the FET in the
saturation regime.
In the same saturation regime, the RSS instability takes place due 
to the transit-time effect of fast-moving electrons in the 
high-field region in the drain side.
Alternatively, the so-called dual-grating-gate structure
(see Fig.~\ref{FigSchematics}(a)),
in which two types of interdigitately-placed gates form
a very efficient grating coupler between THz waves
and 2D plasmons~\cite{Otsuji2006a,Popov2011a}, 
has been proposed for direct THz emission
without antenna integration~\cite{Otsuji2006a,Otsuji2008}.
The RSS instability in this structure
has been investigated analytically~\cite{Ryzhii2008c}.
In addition, asymmetry of the gate placement expects to lead to 
partial realization of the asymmetric boundary conditions 
and, in turn, of the DS 
instability~\cite{Otsuji2014a,Otsuji2015a}.

However, in FETs or high-electron-mobility transistors (HEMTs)
based on usual semiconductors (Si and compound semiconductors 
such as InGaAs and GaN), growth rates of
the instabilities are of the order of $10^{11}$ s$^{-1}$, 
which are limited by electron saturation velocities
($\lesssim 2\times10^{7}$ cm/s in the GaAs channel).
With such low growth rates the plasmons are easily damped out 
at room temperature by a large damping rate 
($\gtrsim 10^{12}$ s$^{-1}$) associated with electron scattering.

Plasmons in graphene have then attracted much attention
owning to its gapless energy
spectrum and massless carriers
~\cite{Ryzhii2006c,Ryzhii2007a,Mikhailov2011,Dubinov2011,
Ju2011,Vicarelli2012,Wang2012a,Svintsov2012,Grigorenko2012,
Popov2013a}.
The most straightfoward yet striking advantage of graphene
plasmons over those in usual semiconductors is
its ultimately low scattering rate at room temperature,
if external scattering sources in graphene such as impurities and 
defects and those induced by the substrate and gate insulator are 
excluded and the electron scattering is limited only 
by the acoustic-phonon scattering~\cite{Hwang2008b} in graphene. 
Then, the plasmon damping rate can be down to $10^{11}$ 
s$^{-1}$~\cite{Satou2013}, together with the electron drift velocity 
up to $\lesssim 10^{8}$ cm/s. These lead to an
expectation that the plasmon instabilities in graphene 
are very strong and can take place at room temperature.
Although the technology of graphene fabrication with
ulitimately high quality still needs to be progressed, 
a recent experimental report on 
graphene encapsulated into hexagonal boron nitride 
layers~\cite{Wang2013a}, which demonstrated the 
electron mobility at room temperature comparable to the 
acoustic-phonon-limited value, supports its feasibility.

In this Letter, we conduct simulations of the plasmon instabilities
in the dual-grating-gate graphene FET with 
dc current injection, assuming the accoustic-phonon-limited
scattering rate at room temperature,
and demonstrate occurence of giant instabilities with their total 
growth rate (which we define as the growth rate subtracting
the damping rate) exceeding $4\times10^{12}$ s$^{-1}$ and with
the plasmon frequency ranging in the THz/mid-infrared range.
We show the gate-length dependence of the plasmon frequency and 
the total growth rate extracted from the simulations, and we
found distinct dependences of growth rates specific to
the DS and RSS instabilities~\cite{Dyakonov1993,Ryzhii2005},
thus identifying the giant total growth rate as simultaneous
occurence, or more specifically, a linear superposition of 
those instabilities.


%

Mechanisms of the DS and RSS instabilities and their features 
are summarized as follows.
The DS instability originates from the Doppler shift effect
at asymmetric boundaries. In case of FETs in the saturation regime,
the potential is fixed at the source side (the short boundary),
while the current, i.e., the electron velocity is fixed
at the drain side (the open boundary). 
Between these asymmetric boundaries, a resonant cavity for plasmons
is formed. Then, the open boundary reflects traveling 
plasmons towards it with the current amplitude preserved,
whereas the short boundary reflects them with amplification.
The resonant frequency of the excited plasmons is determined by 
geometrical factors (gate length, gate dielectric thickness, and 
lengths of ungated regions in the channel) and gate voltage 
through the electron concentration. The latter enables the tuning
of the frequency. On the other hand,
the RSS instability takes place in the same saturation regime due 
to the transit-time effect of fast-moving electrons in the 
high-field region (i.e., the low-concentration region) in the 
drain side. The electric field created by those electrons 
modulates the concentration at the edge of the high-concentration
region and results in the bunching of electrons in the
low-concentration region. Depending on the length of the 
low-concentration region and the electron velocity, 
it interferes with plasmons in the high-concentration region
either constructively or destructively. The constructive
interference corresponds to the instability. The plasmon 
frequency is again determined by the geometrical factors
in the high-concentration region and the gate voltage.

In case of dual-grating-gate structures, both instabilities can be
realized by following configurations.
First, the resonant plasmon cavity can be formed by modulating
the concentration profile through gate voltages in such a way that
high- and low-concentration regions are created.
This also enables the occurence of the RSS instability.
Second, asymmetry of the gate placement leads to 
partial realization of the asymmetric boundary conditions
and, in turn, of the DS instability. 

Figure~\ref{FigSchematics}(a) shows a dual-grating-gate graphene 
FET under consideration. In our simulation model, we assume 
that the number of periods of the grating gates is sufficiently 
large, i.e., the total channel length is longer than the 
THz/mid-infrared wavelength, so that we can ignore effects of 
nonperiodicity such as the presence of the source and drain contacts 
and we can consider one period as a unit cell.
We fix the thickness of the gate dielectric, $W_{\text{g}} = 50$ nm,
the left spacings of the gates 1 and 2,
$s_{\text{g}1} = 200$ nm, $s_{\text{g}2} = 300$ nm, 
whereas we vary the lengths of the gate 1 and 2, $L_{\text{g}1}$
and $L_{\text{g}2}$, in the range between $100$ and $400$ nm
in order to reveal characteristics of the instabilities
obtained in this work by comparison with existing theory.
The values of those parameters were chosen such a way that 
frequencies of self-excited plasmons fall into the THz/mid-infrared
range.
The dielectric constant of media surrounding the graphene channel 
is set to $\epsilon = 4$.
Figure~\ref{FigSchematics}(b) is a profile of the steady-state 
electron concentration without current flow, and it was calculated 
self-consistently 
with a uniform electron doping $\Sigma_{e} = 5\times10^{11}$ 
cm$^{-2}$ together with certain gate voltages.
The slight electron doping is introduced to avoid plasmon damping
due to the electron-hole friction~\cite{Svintsov2012};
even at the point with the lowest electron concentration, the hole
concentration is negligibly low, $\Sigma_{h} = 6.8\times10^{9}$
cm$^{-2}$. On the other hand, the highest electron concentration
is set not so high that the difference between maximum and
minimum Fermi energies ($166$ and $68.5$ meV, respectively) 
does not exceed optical-phonon energies in graphene and thus
electrons injected quasi-ballistically from the region with high
electron concentration do not experience
optical-phonon emission, which would critically hinder 
the RSS instability.

We use the quasi-classical Boltzmann equation to describe 
the electron transport in the channel:
\begin{equation}\label{EqBoltzmann}
  \frac{\p f}{\p t}+\vF\frac{p_{x}}{|\bm{p}|}\frac{\p f}{\p x}
  -eE_{x}\frac{\p f}{\p p_{x}} = J_{\text{LA}}(f|\bm{p}),
\end{equation}
where $v=10^{8}$ cm/s is the Fermi velocity in graphene,
$E_{x}$ is the self-consistent electric field in graphene,
and $\bm{p} = (p_{x}, p_{y})$ is the momentum.
In the right-hand side of Eq.~(\ref{EqBoltzmann}),
we take into account collision integrals for 
the acoustic-phonon scattering, $J_{\text{LA}}$, where
\begin{equation}\label{EqCollisionIntegral}
  J_{\text{LA}}(f|\bm{p}) = \frac{1}{(2\pi\hbar)^{2}}\int d\bm{p}'
  W_{\text{LA}}(\bm{p}'-\bm{p})[f(\bm{p}')-f(\bm{p})]
\end{equation}
and an explicit expression of the transition probability
$W_{\text{LA}}$ can be found in Ref.~\cite{Hwang2008b}.
We use the so-called weighted essentially nonoscillatory
finite-difference scheme~\cite{Carrillo2003b, Galler2006} to
solve Eq.~(\ref{EqBoltzmann}),
which is demonstrated to be applicable for graphene transport 
simulation~\cite{Lichtenberger2011}. The time-step of the
simulation was set to $\Delta t = 0.05$ fs to avoid 
numerical instabilities.
The 2D Poisson equation is solved self-consistently with
Eq.~(\ref{EqBoltzmann}) using a finite-element library called 
libMesh~\cite{libmesh}. Periodic boundary conditions are set
for both the potential and electron distribution function.
More detail of the simulation model is 
described elsewhere~\cite{Satou2013}.

\begin{figure}[t]
  \begin{center}
    \includegraphics[width=8.8cm]{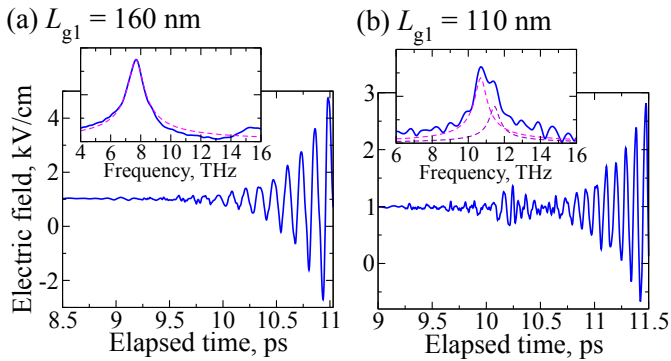}
  \caption{\label{FigTDepField}
    Time evolution of the electric field at the middle of a channel 
    region under the gate 2 with (a) $L_{\text{g}1} = 160$ nm and 
    (b) $110$ nm. In both cases, the external electric field is 
    set to $E_{\text{ext}} = 0.8$ kV/cm.
    The insets show corresponding absolute value of the
    Fourier transform.}
  \end{center}
\end{figure}

A simulation starts by applying a uniform external dc electric field
in the channel direction, $E_{\text{ext}}$, to inject
positive source-drain dc current. 
To avoid unwanted plasmon excitation associated with an abrupt 
turn-on of the dc electric field, an artificially large damping 
factor is set in front of the collision 
integral in Eq.~(\ref{EqBoltzmann}),
and it is gradually decreased to unity within 6 picoseconds,
which is much longer than the inverse of plasmon frequency
obtained in the simulation. Then,
the electron concentration, electric field, etc. 
begin to oscillate and their amplitudes increase with time 
as shown in Fig.~\ref{FigTDepField}(a). This is identified 
as occurence of plasmon instability.

In general, oscillations simulated contain several modes.
With certain sets of parameters distinct fundamental modes
can be obtained, and the frequency and total growth rate can be
easily extracted. However, with other sets they
contain fundamental and second modes with very close frequencies
and/or higher harmonics which have comparable amplitudes with
the fundamental modes, resulting in the beating 
(see Fig.~\ref{FigTDepField}(b)) 
and/or the distortion of waveforms.
To extract the frequency and total growth rate of 
the fundamental mode, 
we perform the Fourier transform of the
oscillation with respect to time, pick up the first peak and 
another close peak, if any, and perform a curve fitting to them
with the following function:
\begin{equation}\label{EqDoubleSqrtLorentzian}
  F(\omega) = \sum_{i=1,2}
  \frac{E_{i}}{4\pi}\frac{|\gamma_{i}|}
  {\sqrt{\gamma_{i}^{2}+(\omega-\omega_{i})^{2}}}.
\end{equation}
Equation~(\ref{EqDoubleSqrtLorentzian}) is equal to 
the absolute value of the Fourier transform of 
the summation of two exponentially growing harmonic functions
$\sum_{i=1,2}E_{i}\exp(\gamma_{i}t)\cos(\omega_{i}t+\theta_{i})$
around its peak(s).
As seen in the insets of Figs.~\ref{FigTDepField}(a) and (b), 
The fundamental and adjacent second modes can be well separated
by this method.


%
\begin{figure}[t]
  \begin{center}
    \includegraphics[width=6.2cm]{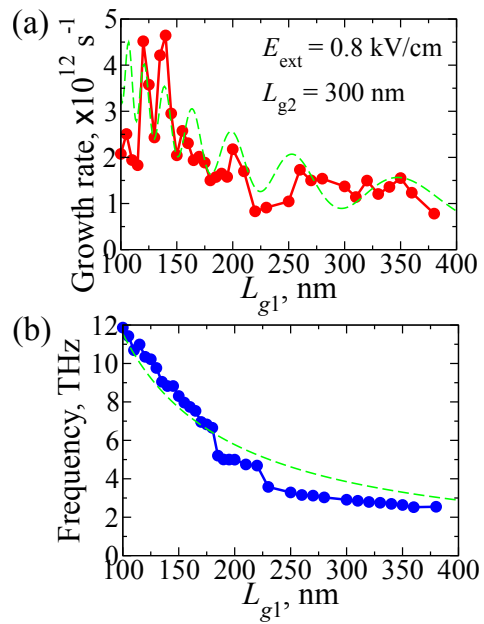}
  \caption{\label{FigFreqGrowthRate}
    (a) Total growth rate and (b) frequency of the fundamental 
    plasmon mode as functions of the length of the gate 1 (circles 
    with solid lines) and theoretical fitting curves 
    (dashed lines).}
  \end{center}
\end{figure}

Figures~\ref{FigFreqGrowthRate}(a) and (b) show the total 
growth rate and frequency of the fundamental mode as 
a function of the length of
the gate 1 with $E_{\text{ext}} = 0.8$ kV/cm and with 
$L_{\text{g}2} = 300$ nm.
They clearly demonstrates the total growth rate of
the instability exceeding $4\times10^{12}$ s$^{-1}$ with the
frequency in the THz/mid-infrared range. 
This value is an order of magnitude
larger than achievable in FETs or HEMTs based on 
usual semiconductors at room temperature.
This giant total growth rate of the plasmon instability in the 
dual-grating-gate graphene FET is attributed to
the lower plasmon damping and also to the larger drift velocity,
both originating from the lower scattering rate limited only
by acoustic phonons. This point shall be discussed in more detail
shortly later.

\begin{figure}[t]
  \begin{center}
    \includegraphics[width=8.3cm]{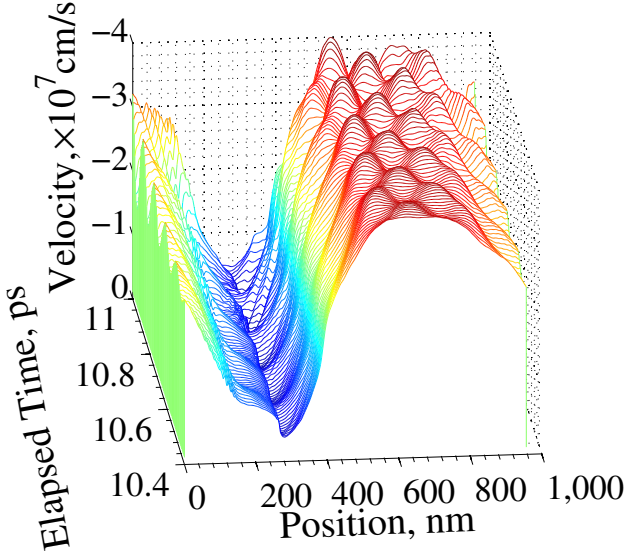}
  \caption{\label{FigPosDepVelocity}
    Time evolution of the electron velocity distribution
    after the onset of plasmon instability, in the case with 
    $L_{\text{g}1} = 160$ nm and $L_{\text{g}2} = 300$ nm.}
  \end{center}
\end{figure}

The gate-length dependence of the total growth rate
exhibits an oscillatory behavior; the oscillation period becomes
shorter and the amplitude becomes larger as the gate length
becomes shorter.
It is specific to the RSS instability, and it corresponds 
to the constructive/destructive interference between
plasmons in the high-concentration region (under the gate 1) and 
the bunched electrons in the low-concentration region (under the 
gate 2).
This effect is illustrated in Fig.~\ref{FigPosDepVelocity} 
where an oscillation of the electron velocity
is built up at the left edge of the high-concentration region
(near $x = 150$ nm), where the electrons are injected to
the low-concentration region.
Besides, there is a monotonically increasing portion of
the total growth rate
with decreasing $L_{\text{g}1}$ in Fig.~\ref{FigFreqGrowthRate}(a).
This can be attributed to the DS instability, in which the
growth rate is inversely porportional to the traveling time
of plasmons in the gated region and therefore, to the
gate length~\cite{Dyakonov1993}.
Those characteristics signify the simultaneous occurence of
the DS and RSS instabilities. In fact, the overall characteristics of
the growth rate can be described qualitatively well by the formula
according to Ref.~\cite{Ryzhii2006e}, 
which is a linear superposition of the growth rates of
those instabilities:
\begin{equation}\label{EqGrowthRate}
  \gamma = \frac{v_{_\text{DS}}}{L_{\text{g}1}}
  -\frac{v_{_\text{RSS}}}{L_{\text{g}1}}
  J_{0}\left(\frac{L_{\gamma}}{L_{\text{g}1}}\right),
\end{equation}
where $v_{_\text{DS}} = 4\times10^{7}$ cm/s, 
$v_{_\text{RSS}} = 7.5\times10^{7}$ cm/s, and 
$L_{\gamma} = 5800$ nm are fitting parameters
(see the fitting curve in Fig.~\ref{FigFreqGrowthRate}(a)).

As seen in Fig.~\ref{FigFreqGrowthRate}(b), 
the frequency is almost inversely proportional
to $L_{\text{g}1}$, and it obeys the
well-known dispersion of gated 2D plasmons,
\begin{equation}\label{EqFreq}
  f = \frac{s_{\text{g}1}}{L_{\text{g}1}},
\end{equation}
where $s_{\text{g}1} = 1.15\times10^{8}$ cm/s is the plasmon phase 
velocity which is extracted as a fitting parameter and
which is quantitatively consistent with the value calculated 
analytically~\cite{Ryzhii2006c} for the region under the gate 1.

\begin{figure}[t]
  \begin{center}
    \includegraphics[width=6.2cm]{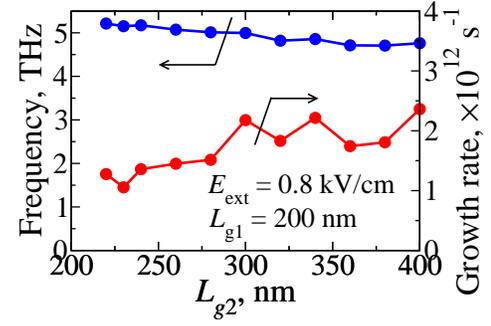}
  \caption{\label{FigFreqGrowthRateLg2Dep}
    (a) Total growth rate and frequency of the fundamental 
    mode as functions of the length of the gate 2.}
  \end{center}
\end{figure}

The giant instability found here at room temperature originates 
from the ultimately weak electron scattering rate in graphene, 
and there are two factors for this:
(1) the weak damping of plasmons and 
(2) the large drift velocity in both high- and 
low-concentration regions that leads to large DS and RSS
instabilities, respectively. First, the estimated
damping rate was around $1.2\times10^{11}$ s$^{-1}$ in the 
low-concentration region and $2.5\times10^{11}$ s$^{-1}$
in the high-concentration region; note that the electron 
scattering rate, and thus the plasmon damping rate, for
the acoustic-phonon scattering is proportional
to the square root of the electron 
concentration~\cite{Hwang2008b,Satou2013}.
Such low values of the plasmon damping rate 
can be achieved only at nitrogen temperature or lower
in other materials.
It is worth mentioning that 
the low-concentration region with adjacent ungated regions 
acts as a better (passive) plasmon resonant cavity,
while the instabilities take place primarily
for plasmons in the high-concentration region, as is evident
from the dependence of the frequency on $L_{\text{g}1}$ 
and its consistency with Eq.~(\ref{EqFreq}),
and the frequency is insensitve to the length of the gate 2
as shown in Fig.~\ref{FigFreqGrowthRateLg2Dep}
(in contrast, the total growth rate can vary with it).
This is similar to the situation 
discussed in Ref.~\cite{Popov2008a},
where the ungated plasmon resonance can be effectively tuned
by the gated region of the channel.
In fact, it can be seen in Fig.~\ref{FigPosDepVelocity} 
that a higher harmonic oscillation with more than one nodes is 
excited in the low-concentration region. This together with the 
oscillation in the high-concentration region form the fundamental 
mode of the whole period.

In addition,
values of drift velocities obtained in the simulation,
$\simeq 4\times10^{7}$ cm/s in the low-concentration region
and $\simeq 10^{7}$ cm/s in the high-concentration region,
confirm the factor (2) for the giant growth rate.
Especially the former is more than twice larger than the
saturation velocity in Si and InGaAs channels. However, 
the velocities extracted by Eq.~(\ref{EqGrowthRate}) as 
fitting parameters differ from these values by several factors.
A reasonable explanation to this discrepancy is that 
the DS instability also takes place not only in the 
high-concentration region but also in the low-concentration
region.
Fig.~\ref{FigPosDepVelocity} exhibits waves propagating in 
opposite directions in the latter, suggesting the occurence
of the DS instability there. This should add a constant term
in Eq.~(\ref{EqGrowthRate}), and then the fitting velocities
should become closer to the simulated values.


In conclusion, we have conducted simulations of plasmon instability
driven by dc current injection in the 
dual-grating-gate graphene FET. We have obtained a giant 
total growth rate of the instability at room temperature 
exceeding $4\times10^{12}$ s$^{-1}$.
Through the dependences of the total growth rate and frequency 
on the gate lengths, we have revealed that the giant total 
growth rate originates from simulateneous occurence of the DS 
and RSS instabilities.
The result obtained strongly suggests that a graphene FET
with the dual-grating-gate structure is very promising for
the realization of a high-power, compact, room-temperature 
operating THz source.


This work was financially supported by JSPS Grant-in-Aid for 
Young Researcher (\#26820122), by JSPS Grant-in-Aid for Specially 
Promoted Research (\#23000008),
and by JSPS and RFBR under Japan-Russia Research Cooperative 
Program. The simulation was carried out using the computational 
resources provided by Research Institute for Information Technology 
Center, Nagoya University, through the HPCI System Research Project 
(\#hp140086), by the Information Technology Center, 
the University of Tokyo, and by Research Institute for 
Information Technology, Kyushu University.

\bibliography{bibdata}

\end{document}